\documentclass[letterpaper, 11pt]{article}
\usepackage[utf8]{inputenc}
\usepackage[in]{fullpage}
\usepackage{setspace}
\usepackage{amsmath, amsfonts, amssymb, bm}

\usepackage[style=nature, block=ragged]{biblatex}
\usepackage[colorlinks=true,
            linkcolor=gray,
            urlcolor=gray,
            citecolor=gray]{hyperref}

\usepackage{color}
\usepackage[usenames,dvipsnames,svgnames,table]{xcolor}
\usepackage{graphicx}
\usepackage{forloop}

\usepackage{appendix}

\title{Using Satellite Imagery and Deep Learning to Evaluate the Impact of Anti-Poverty Programs}
\author{
    Luna Yue Huang%
    \footnote{
        Correspondence: \texttt{yue\_huang@berkeley.edu}.
        $^1$Department of Agricultural and Resource Economics, UC Berkeley, Berkeley, CA, USA.
        $^2$Global Policy Laboratory, Goldman School of Public Policy, UC Berkeley, Berkeley, CA, USA.
        $^3$National Bureau of Economic Research, Cambridge, MA, USA.
    }
    $^{,1,2}$, Solomon Hsiang$^{2,3}$, Marco Gonzalez-Navarro$^{1}$
}
\date{}

\begin{document}

\maketitle

\begin{center}
    This Version: \today
    \vspace{2em}
\end{center}
\vspace{2em}

\begin{abstract}
    The rigorous evaluation of anti-poverty programs is key to the fight against global poverty. Traditional evaluation approaches rely heavily on repeated in-person field surveys to measure changes in economic well-being and thus program effects. However, this is known to be costly, time-consuming, and often logistically challenging. Here we provide the first evidence that we can conduct such program evaluations based solely on high-resolution satellite imagery and deep learning methods. Our application estimates changes in household welfare in the context of a recent anti-poverty program in rural Kenya. The approach we use is based on a large literature documenting a reliable relationship between housing quality and household wealth. We infer changes in household wealth based on satellite-derived changes in housing quality and obtain consistent results with the traditional field-survey based approach. Our approach can be used to obtain inexpensive and timely insights on program effectiveness in international development programs.

\end{abstract}
\clearpage

\begin{refsection}

\section{Introduction}

Rigorous impact evaluation forms the basis of the modern approach to fight global poverty and provides input for evidence-based policy making \cite{deaton1997analysis, banerjee2011poor}. The impacts of anti-poverty interventions are almost universally evaluated using household surveys, typically comprehensive questionnaires containing hundreds of questions that can touch every aspect of people's lives. However, such field surveys are often prohibitively expensive to conduct \cite{pamies2015development, alix2021better} and unanticipated events, such as political unrest or public health crises, frequently disrupt them \cite{brune2020social}. In this paper, we provide the first demonstration that the household welfare impacts of a large scale anti-poverty randomized controlled trial (RCT) can be accurately measured relying solely on satellite data, instead of household surveys.

Recent advances enable poverty to be identified remotely \cite{jean2016combining, blumenstock2016fighting, engstrom2017poverty, babenko2017poverty, watmough2019socioecologically, yeh2020using, aiken2020targeting, blumenstock2020machine} and widespread adoption of mobile phones allows targeted anti-poverty interventions to be deployed over-the-network \cite{suri2017mobile}, such as the cash transfer program we study here \cite{egger2019general}. By demonstrating that the impacts of such interventions can be evaluated remotely, we hope that future programs can, in principle, be designed, deployed, and evaluated with limited reliance on logistically complex and expensive ground operations. Because costs and logistics play a major role in limiting the scale of anti-poverty programs \cite{blattman2014show}, simplifying their deployment and evaluation is crucial to achieving their full global potential.

We study a large pre-existing trial \cite{egger2019general} that was recently completed and evaluated with field surveys and show that we can consistently recover the impact of the program using satellite imagery and deep learning. While previous studies have successfully evaluated the environmental impacts of randomized controlled trials with remote sensing data \cite{alix2013ecological, jayachandran2017cash}, we are not aware of studies that demonstrate similar successes for household economic well-being. Specifically, we combine high-resolution daytime imagery \cite{gsm} and state-of-the-art deep learning models \cite{he2017mask} to measure housing quality among treatment and control households, and estimate the program effects on housing quality. We then map housing quality to household wealth for these households by inverting an ``Engel curve,'' an established concept in economics \cite{elbers2003micro, tarozzi2009using, young2012african, atkin2020new} that describes household spending on specific goods as a function of economic well-being. Using this approach, we accurately recover the program effects on household wealth for a fraction of the cost (\$0.006 per household, see Supplementary Materials \ref{sec:appendix-cost}) that would typically be spent on household surveys (\$18--300 per household \cite{alix2021better}).

Early work has shown that satellite data can be used to monitor economic development by correlating nighttime luminosity, i.e., the amount of light emitted from Earth at night (hereafter ``night light'') with Gross Domestic Product (GDP) at national and subnational scales \cite{henderson2012measuring, chen2011using, michalopoulos2014national}. However, the night light data show poor sensitivity in less developed and rural areas \cite{jean2016combining}, presumably because of low electrification rates---for example, from 1992 to 2008, 99.73\% of pixels were completely unlit in Madagascar, 99.47\% in Mozambique, and this is representative of low-income countries \cite{henderson2012measuring}. This makes the data less useful for studying the very target of many international development programs---people living under the poverty line. Additionally, the low spatial granularity of night light prevents it from being used to evaluate programs reliant on fine spatial variations, including most randomized controlled trials in which households in close proximity to one another are assigned to different treatments.

We propose an alternative approach---we analyze daytime imagery using a deep-learning model \cite{he2017mask} to explicitly measure the quality of housing, a tangible and verifiable asset that is known to be a powerful proxy for household wealth. Even in communities where electrification rates are low, housing quality remains a strong predictor of wealth, in part because housing accounts for a sizable portion (10--20\%) of total household expenditure globally \cite{oecd2014}. Furthermore, in many rural and low-income contexts, individuals do not migrate often \cite{murrugarra2010migration} and tend to frequently upgrade their housing by expanding or building new structures on their property, making housing footprint a meaningful proxy for welfare that responds to improved economic conditions. In this study, we focus on building footprint because it can be precisely measured at scale with modern deep learning techniques.

Many features of buildings other than footprint are observable with satellite imagery; for example roof material \cite{marx2019there, michaels2017planning}. One of the main advantages of the method proposed here compared to alternative ``black-box'' machine learning approaches to measuring wealth that utilize \emph{all} available information contained in satellite images (such as convolutional neural networks \cite{jean2016combining, yeh2020using} or random kitchen sinks \cite{rolf2020generalizable}) is that it allows the exclusion of subsets of satellite-derived outcomes that may have been directly impacted by the intervention. We show the benefits of this feature of our method in the context of the experiment we evaluate. Specifically, households were eligible for the GiveDirectly study as long as their roofing was of low quality (thatched). Due to this eligibility criterion, treatment households were ``prompted'' to use the GiveDirectly transfer to upgrade their roofing as a way to signal to the experimenters that they had used the cash for good. An improvement of roofs among participating households beyond what would have been expected solely from wealth increases biases estimates of wealth when methods cannot exclude subsets of outcomes. In contrast, it is straightforward for our method to focus exclusively on subsets of available information that were not affected directly (in this case building footprints) while ignoring problematic outcomes (such as roof material) in order to  provide unbiased estimates of wealth effects.
\section{Results}

We evaluate a development intervention that was conducted in 2014--2017 in 653 villages in rural Kenya \cite{egger2019general}. GiveDirectly, a US charity, implemented a randomized controlled trial of unconditional cash transfers to rural households via mobile money, using as sole eligibility criterion whether the household lived under a thatched roof (a low quality roof material that served as a simple means test). Each treatment household received \$1,000---equivalent to about 75\% of annual household expenditures---in lump sum, and could spend it however they wished. To evaluate the effectiveness of the program, GiveDirectly randomly selected 328 villages as the treatment group, where eligible households (about 1/3 of the population) received transfers, and used the remaining 325 villages as the control group. The authors conducted extensive household surveys before and after the distribution of the transfers to measure program impacts as is the current practice in the evaluation literature.

\paragraph{Mapping Treatment Intensity and Housing Quality.}

To evaluate program impacts, we first construct a map that shows the intensity of the anti-poverty program (hereafter ``treatment'') in different geographical units (in this case it is simplest to work with raster grid cells). This geocoded information is obtained from program implementation records, which document where the program was administered. Because of the extremely high granularity of satellite-derived housing quality metrics, it is feasible to study programs that induce fine spatial variation such as household-level randomized trials. Importantly, the variation in treatment intensity has to be either random (if induced by an experiment) or as good as random (in a natural experiment setting), as is the case for any credible program evaluation project.

\begin{figure}[!ht]
\centering
\includegraphics[width=6.5in]{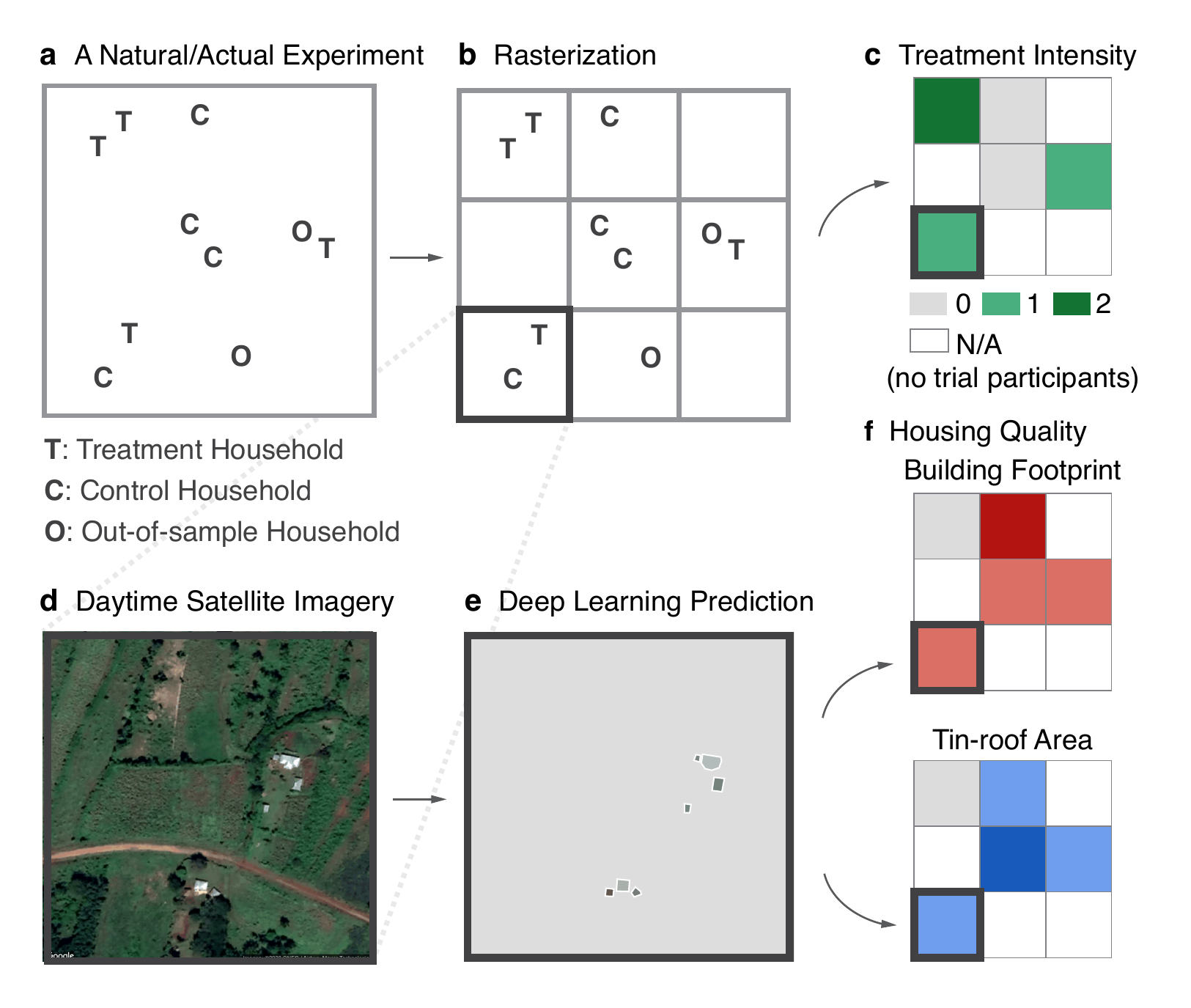}
\vspace{-1em}
\caption{
    \textbf{Constructing maps of treatment intensity and remotely sensed outcomes from program implementation records and satellite imagery.}
    \textbf{a} An illustration of geocoded program implementation records.
    \textbf{b} Placing a regular grid over \textbf{a} and measuring the intensity of the treatment in each grid cell.
    \textbf{c} Constructed raster of the number of treatment households in each grid cell.
    \textbf{d} An example daytime satellite image from Google Static Maps.
    \textbf{e} Example deep learning predictions on \textbf{d}. Each building is outlined in white and filled with the ``representative'' roof color.
    \textbf{f} Constructed rasters of remotely sensed housing quality outcomes.
    In \textbf{c} and \textbf{f}, grid cells without trial participants are omitted and shown in white.
}
\label{fig:schematic}
\end{figure}

For the GiveDirectly experiment, we construct the treatment intensity map from a local census fielded in 2014--2015, which surveyed all the 65,385 households living in the study area \cite{egger2019general}. The census data record each household's geo-location, and indicate whether they belong to the treatment (T), control (C), or out-of-sample (O) group (Figure \ref{fig:schematic}a). Among the three groups, only the treatment households eventually received the cash transfer from GiveDirectly. The control households were randomized into not receiving the transfer, whereas the out-of-sample households were never eligible to participate in the program. Our sample contains 11,055 treatment households and 10,682 control households in total. We lay out a regular grid, and count the number of treatment households in each grid cell (Figure \ref{fig:schematic}b). As every transfer was roughly USD 1,000, this variable can be interpreted as the amount of cash infusion (in \$1,000) into a given grid cell, and is our preferred measure of treatment intensity (Figure \ref{fig:schematic}c).

Next, we measure housing quality in daytime satellite images with deep learning techniques. The input images are from Google Static Maps \cite{gsm}. They are taken after the GiveDirectly intervention, have a spatial resolution of about 30cm per pixel, and contain only the RGB (red, green, blue) bands (Figure \ref{fig:schematic}d).

To segment buildings, we train a state-of-the-art deep learning model, Mask R-CNN \cite{he2017mask}, on large, publicly available datasets such as COCO (Common Objects in Context) \cite{coco} and Open AI Tanzania \cite{openaitanzania}, as well as a small annotated dataset, which are randomly sampled from all the input images (see Supplementary Materials \ref{sec:appendix-train} for details on model training). The model predictions are highly accurate, both quantitatively (Supplementary Figure \ref{fig:prcurve}) and qualitatively (Supplementary Figure \ref{fig:chips}). The model generalizes well to other countries, such as Mexico, where the number of houses identified in the deep learning predictions is highly correlated with the census population count (Supplementary Figure \ref{fig:mx} and Supplementary Materials \ref{sec:appendix-mx}). After post-processing, each predicted instance of buildings is represented by a polygon and a ``representative'' roof color (Figure \ref{fig:schematic}e). The Mask R-CNN model conducts instance segmentation (as opposed to semantic segmentation), meaning that it is able to identify every building instance separately, even if they are adjacent to each other. As such, we can measure housing outcomes for each household.

We extract two metrics for each built structure: the size of building footprint, and the type of roof material. The roofs are classified into three types: tin roof, thatched roof, and painted roof, based on their color profiles (Supplementary Figure \ref{fig:colors}). Compared to tin roofs, thatched roofs are generally of lower quality \cite{haushofer2016short, egger2019general}. (Painted roofs are relatively uncommon in the study area.) In prior work, roof reflectance and roof color have been shown to be good proxies of housing quality \cite{marx2019there, michaels2017planning}. As such, we aggregate the total building footprint to measure all housing assets (Figure \ref{fig:schematic}f, Building Footprint), and the footprint of tin-roof buildings to measure high-quality housing assets (Figure \ref{fig:schematic}f, Tin-roof Area), in each grid cell. To obtain night light data for systematic comparison, we download and resample the Visible Infrared Imaging Radiometer Suite (VIIRS) Day/Night Band (DNB) composite images in 2019 \cite{viirs, elvidge2017viirs}.

\begin{figure}[!ht]
\centering
\includegraphics[width=6.5in]{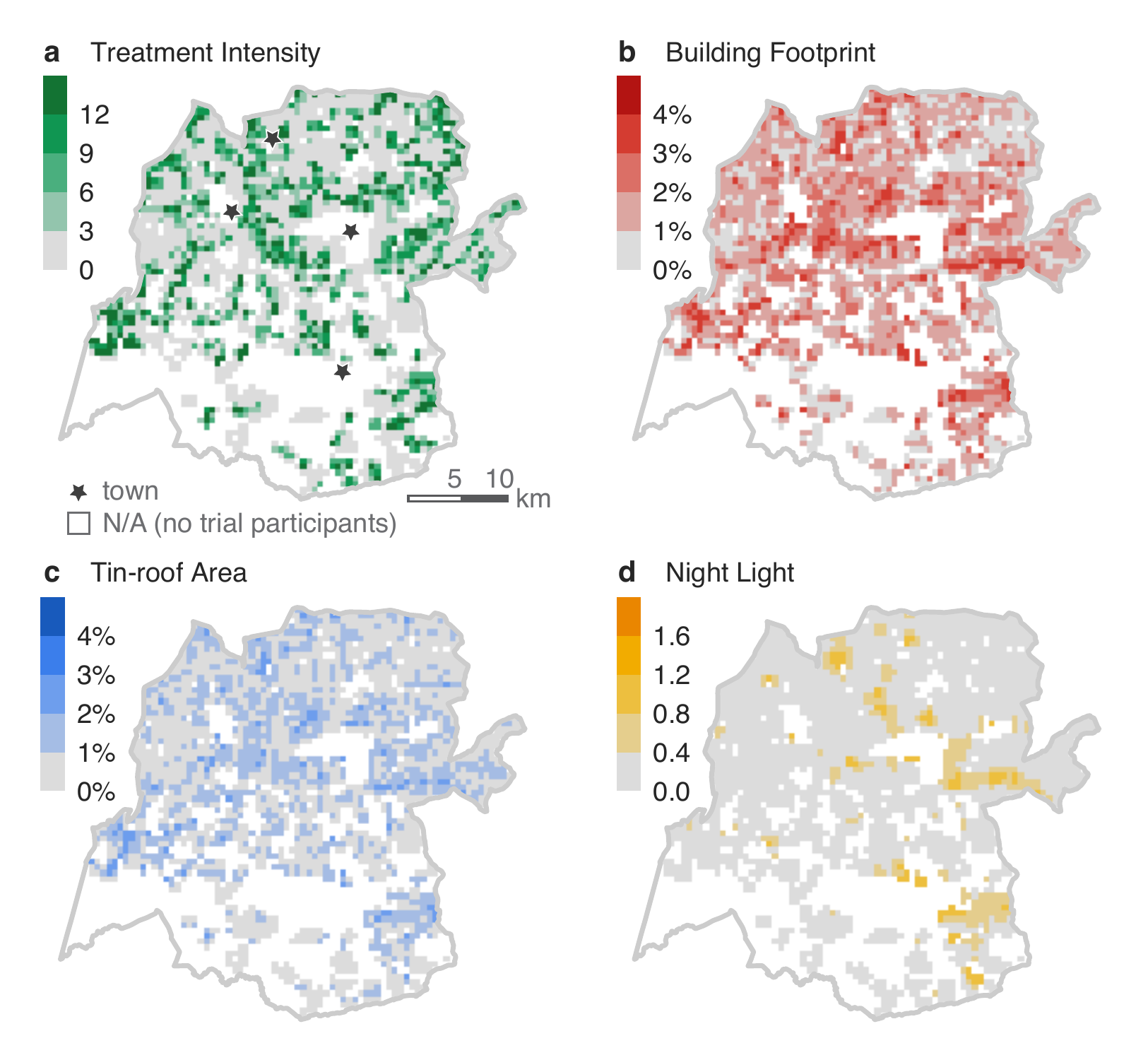}
\caption{
    \textbf{Mapping treatment intensity and remotely sensed outcomes in the GiveDirectly study area in 2019.}
    \textbf{a} Treatment intensity represents the number of households who received a \$1,000 cash transfer from GiveDirectly.
    \textbf{b} Building footprint measures the total area covered by any building, shown as a percentage of the total area.
    \textbf{c} Tin-roof area measures the total footprint of buildings with roofs made of tin (a high quality construction material), shown as a percentage of the total area.
    \textbf{d} Night light is the average radiance in the Visible Infrared Imaging Radiometer Suite (VIIRS) Day/Night Band (DNB).
    In all the panels, the gray lines outline the GiveDirectly study area in Siaya, Kenya. Grid cells without trial participants are omitted and shown in white. $n=2,501$.
}
\label{fig:map}
\end{figure}

The maps of treatment intensity and remotely sensed outcomes for the GiveDirectly experiment are shown in Figure \ref{fig:map}. For visual display and privacy protection purposes, we plot the maps with a spatial resolution of $0.005^{\circ}$ (roughly 500 meters), which is lower than the resolution used in the subsequent statistical analysis. The experiment generated substantial variation in treatment intensity, as expected (Figure \ref{fig:map}a). Both of the housing quality measures capture richer variation in the entire area (Figure \ref{fig:map}b, c), whereas the night light data demonstrate little variation in this rural, sparsely populated area, except in a few spots close to local towns (Figure \ref{fig:map}d).

\paragraph{Estimating the Program Effects on Housing Quality.}

\begin{figure}[!ht]
\centering
\includegraphics[width=4in]{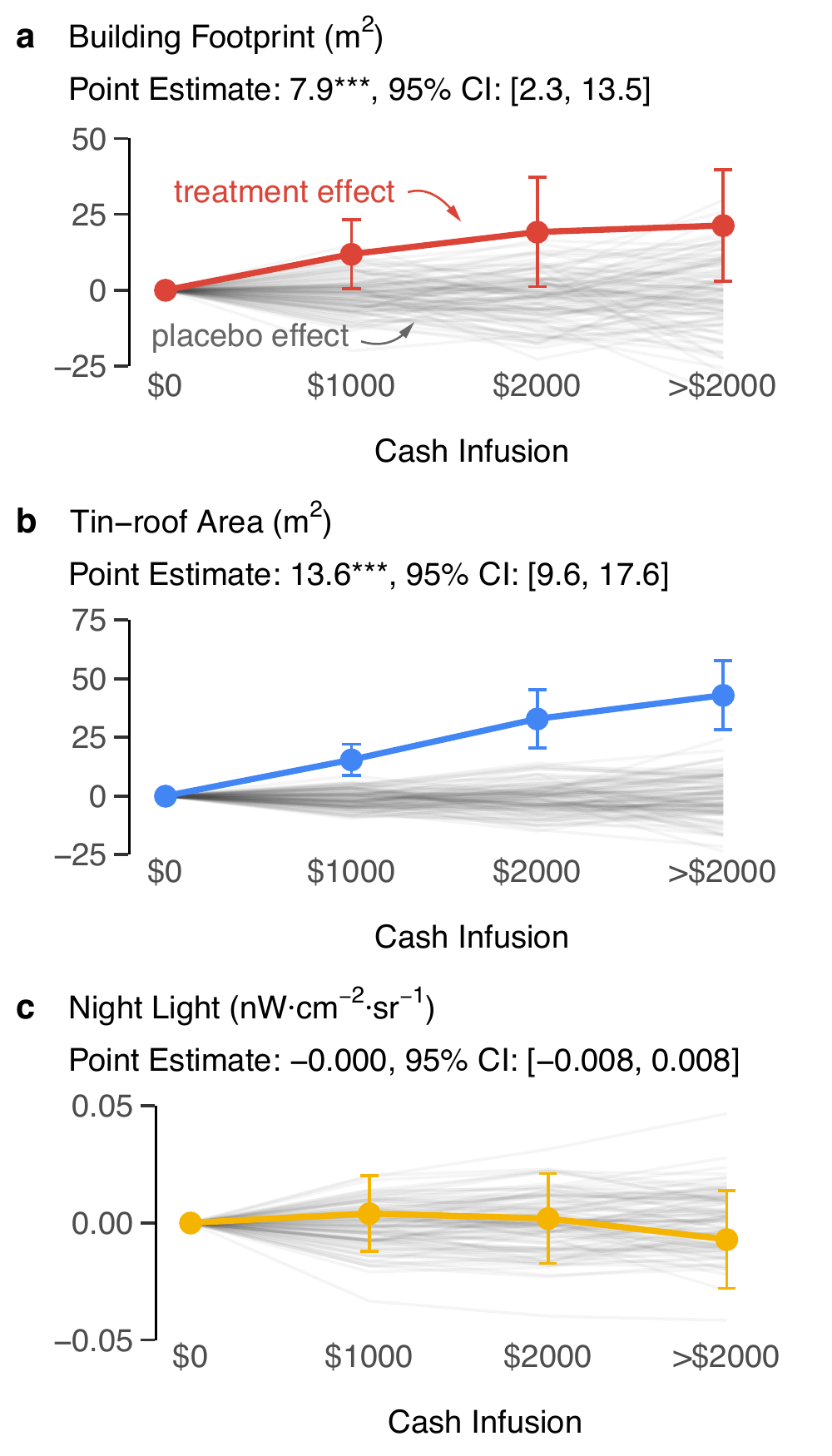}
\caption{
    \textbf{Housing quality increased in response to the GiveDirectly cash transfer, but night light remained unchanged.}
    The treatment effects of the cash transfers on building footprint (\textbf{a}), tin-roof area (\textbf{b}), and night light (\textbf{c}) are shown in color. The dots represent the point estimates, and the error bars represent the 95\% confidence intervals.
    Gray lines show the estimated effects of the placebo cash infusions from 100 simulations.
    The panel subtitles report the average treatment effect of a \$1,000 transfer and the 95\% confidence intervals, assuming constant effect. *** indicates statistical significance at the 1\% level for a two-sided t-test. $n=14,155$.
}
\label{fig:ate}
\end{figure}

We regress the remotely sensed outcomes on treatment intensity to estimate the causal effects of the GiveDirectly cash transfer. We choose a spatial resolution of $0.001^{\circ}$ (approximately 100m), such that most of the grid cells contain 0--5 households. We exploit only the experimentally-induced random variation in treatment intensity for identification, and account for pre-determined differences in program eligibility. Intuitively, consider two grid cells, one containing a household that received the transfer, and the other containing a household that was eligible to get the transfer but did not because it was randomized into the control group. With valid randomization \cite{egger2019general}, the differences in outcomes between the two can be attributed to the cash transfer. We plot the causal effects on night light and housing quality as cash infusion intensity increases (Figure \ref{fig:ate}, in color), without making assumptions on the structure of the effects. The results suggest that the effects grow linearly with the amount of cash infusion. We therefore also report an ``average'' effect, estimated with the assumption that each \$1,000 transfer generates an effect of the same magnitude (Figure \ref{fig:ate}, panel subtitles). We demonstrate the validity of the empirical strategy further by running 100 placebo simulations---we artificially generate placebo cash transfers that did not actually take place but is consistent with the original randomization design, and estimate their treatment effects (Figure \ref{fig:ate}, in gray). The resulting estimates are reassuringly centered around zero.

We observe statistically significant and economically sizable effects on housing quality, on both the extensive margin (larger building footprint) (Figure \ref{fig:ate}a), and the intensive margin (higher quality roofs) (Figure \ref{fig:ate}b). On average, a \$1,000 cash transfer significantly increased building footprint by 7.9 square meters (95\% CI: [2.3, 13.5], $t(14,143)=2.8$, $p=0.006$) or 85.0 square feet, and tin-roof area by 13.6 square meters (95\% CI: [9.6, 17.6], $t(14,143)=6.7$, $p<0.001$) or 146.4 square feet. These increases indicate that households may have built new structures---either primary residences or auxiliary structures, such as kitchens and sheds, expanded their existing structures, and/or upgraded their thatched roofs to tin roofs, an improvement that people commonly used the transfer for \cite{haushofer2016short}. These estimates are consistent with the results from extensive field surveys, which also documented large increases in housing asset values \cite{egger2019general}.

On the other hand, we do not observe any program effects on night light (Figure \ref{fig:ate}c), despite the fact that the cash transfer had large positive impacts on many aspects of the recipient households' economic well-being---food expenditure, consumer durable spending, asset holding, and housing values \cite{egger2019general}. The estimated effect is $-0.000120$ (95\% CI: [$-$0.008, 0.008], $t(14,143)=-0.03$, $p=0.977$) which is not statistically different from zero, is small in magnitude, and actually slightly negative. This may be because of low demand for electrification \cite{lee2020experimental}, or the poor sensitivity of night light in low-income, rural regions \cite{jean2016combining}.

\paragraph{Recovering the Program Effects on Economic Well-being with Engel Curves.}

\begin{figure}[!ht]
\centering
\includegraphics[width=6in]{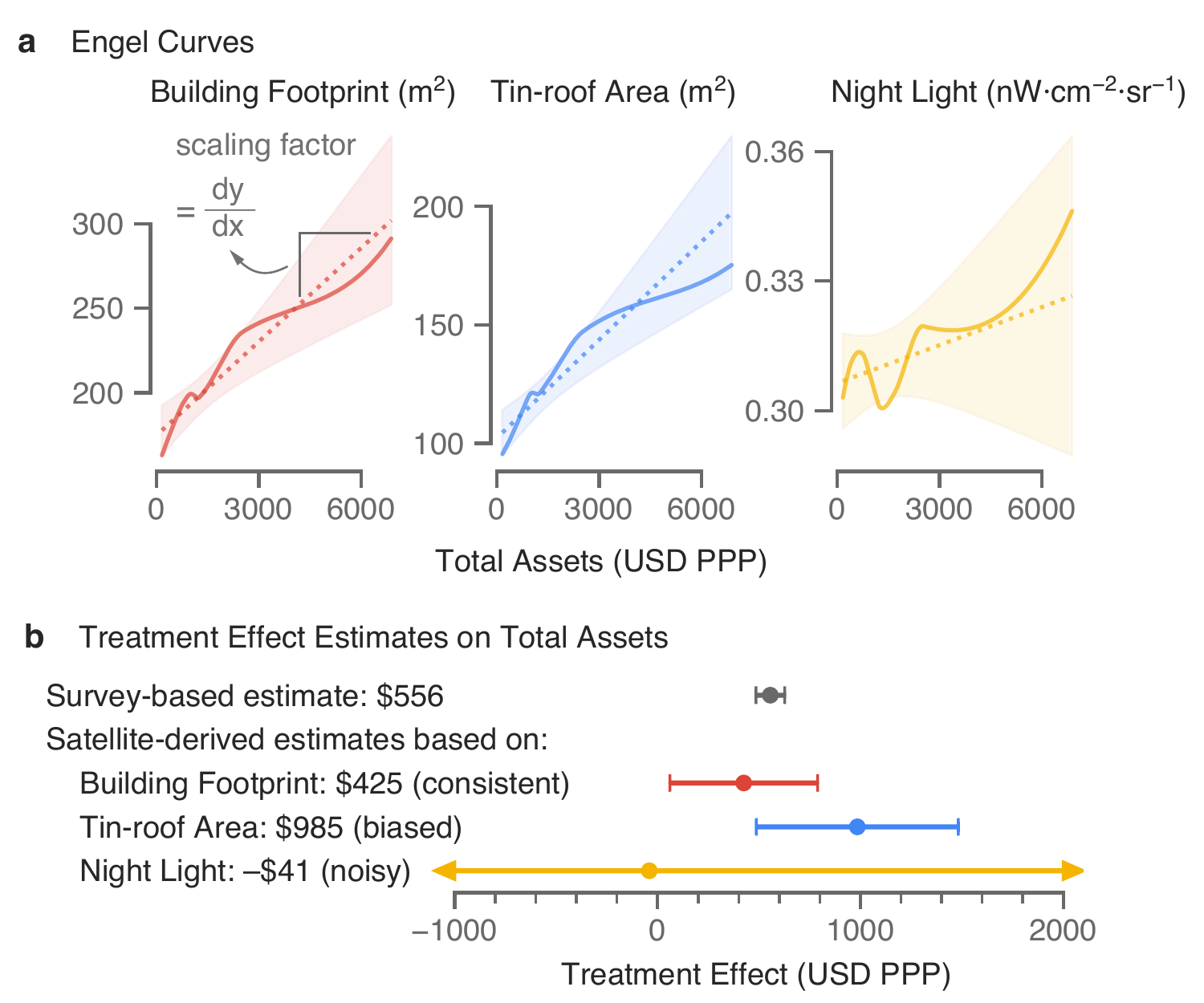}
\caption{
    \textbf{The treatment effect on total assets can be correctly recovered by scaling the effect on building footprint.}
    \textbf{a} The Engel curves of building footprint, tin-roof area, and night light, estimated with LOESS (solid line) or a linear regression (dotted line). The shaded regions represent the 95\% confidence intervals for the latter.
    \textbf{b} Comparing the survey-based versus satellite-derived treatment effects. The dots show the point estimates. The error bars show the 95\% confidence intervals, with the arrow(s) marking upper/lower bounds that are out of range (if any).
    $n=1,844$.
}
\label{fig:engel}
\end{figure}

We recover the program effects on household economic well-being with a canonical economic concept, the Engel curve. Engel curves describe how household expenditures on particular goods or services depend on households' economic well-being. For example, it is widely known that poorer families spend a larger share of their expenditure on food. Engel curves have long been used to infer economic well-being without needing detailed information on prices as it is straightforward to measure how much of a household's expenditure is spent on food \cite{elbers2003micro, tarozzi2009using, young2012african, atkin2020new}. We adapt this concept to housing quality by exploiting the fact that someone who lives in a larger house is likely to be wealthier than someone who lives in a smaller house (Figure \ref{fig:engel}a). By the same logic, if we observe that someone's house size increased, then we can  infer what level of wealth is associated to such a house size---as if they were moving up on the Engel curve. Mathematically, the slope of the Engel curve represents the ratio between the change in house size and the change in wealth. We divide the change in the house size (Figure \ref{fig:ate}) by the slope of the Engel curve (Figure \ref{fig:engel}a) to infer the corresponding change in wealth (Figure \ref{fig:engel}b). Importantly, the validity of this approach depends on the assumption that the Engel curve does not shift in response to the treatment, which could happen due to relative price changes of the good or taste changes.

In this study, we derive housing Engel curves from an endline survey of the GiveDirectly trial participants between May 2016 and June 2017, which includes 4,578 geo-coded households who were eligible for the transfer. Of these households, only those assigned to the control group are used for the estimation. In Figure \ref{fig:engel}a, we show the relationship between survey-based measures of economic well-being ($x$-axis) and remotely sensed night light or housing quality measures ($y$-axis). The Engel curves are estimated with a linear regression (dotted lines). The non-linear fit with LOESS (solid lines) shows only small deviations from the linear regression line, and we cannot reject the null hypothesis that these Engel curves are linear (see Methods). The Engel curves are also roughly monotonically increasing, validating the choice of these variables as wealth proxies.

The Engel curves can be derived from any geo-coded consumption and expenditure survey, as long as the surveyed households are---or can be re-weighted to be---representative of the sample in the previous treatment effect estimation step. Notably, the sample does not necessarily have to include any one who has received the treatment, opening up the possibilities of using existing data sources (such as the Living Standards Measurement Study (LSMS)) to estimate Engel curves. We demonstrate this by comparing the Engel curves derived from two distinct samples: the households who were deemed eligible to receive the cash transfers (meaning that they used to live in thatched-roof houses), and households who were not. While all the households live in the same area in western Kenya, the ineligible households are generally wealthier than the eligible ones. Their Engel curves, however, are similar within the same range of wealth (Supplementary Figure \ref{fig:engel-ei-diff}).

We scale program effects on each remotely sensed outcome by the Engel curve slope to estimate the impacts of the GiveDirectly transfer on household wealth, measured by aggregating the values of a variety of assets as measured with household surveys. In Figure \ref{fig:engel}b, we compare the satellite-derived estimates against the survey-based estimates, which are computed from rich endline household survey data and taken from Table 1, Column 1 in the original paper \cite{egger2019general}. As can be seen, the estimate based on building footprint (USD 425 PPP, 95\% CI: [61, 788]) is informative and very close to the survey based estimate (USD 556 PPP, 95\% CI: [485, 626]). For reference, the entire GiveDirectly cash transfer is worth USD 1,871 PPP (USD 1,000 nominal). Note that the estimate based on night light is slightly negative and imprecise, and both the upper and lower bounds are uninformative. In contrast, the estimate based on tin-roof area is about two times as large as the survey-based estimate. The results are qualitatively similar when we distinguish between housing asset (Supplementary Figure \ref{fig:engel-housing}) and non-housing asset (Supplementary Figure \ref{fig:engel-nonhousing}), or when we use annual consumption expenditure as the alternative measure of economic well-being (Supplementary Figure \ref{fig:engel-consumption}).

Why is the estimate based on the tin-roof area much larger than the survey based estimate? We argue this is due to the violation of a key assumption, which is that the Engel curve used to estimate changes in wealth cannot change directly in response to the treatment---only through its wealth effects. To give intuition for why this matters, consider a  program that directly gives people food. In such a case we can no longer look at food consumption to infer  program effects on economic well-being, because the relationship between the food and income will be altered directly by the program and households will ``look'' wealthier than they really are based on their food consumption. More relevant for impact evaluation using satellite data, this example is analogous to examining the impacts a program that provides roads to a region. One would need to exclude the program roads themselves contained in satellite images and look at other correlates of welfare to estimate impacts of such a roads program in an unbiased manner. In the GiveDirectly case, only households that lived in thatched-roof houses were eligible for the study. Households' usual consumption patterns of high-quality tin roofs might have been affected by this eligibility criteria. One can observe that treatment households owned more tin-roof buildings compared to control households with the same amount of wealth (Supplementary Figure \ref{fig:engel-tc-diff}). This may have been a result of households interpreting the treatment as a ``labelled'' cash transfer \cite{benhassine2015turning}.

These results highlight the importance of using interpretable proxies when evaluating programs with machine learning predictions. An emerging literature is making great progress in mapping poverty with satellite imagery and machine learning with a high spatial granularity at scale \cite{jean2016combining, yeh2020using, blumenstock2016fighting, watmough2019socioecologically, aiken2020targeting, blumenstock2020machine, engstrom2017poverty, babenko2017poverty}. Typically, a machine learning model first learns the mapping between the input satellite images and the ground truth labels of wealth or consumption expenditure, assembled from geo-coded household surveys. Then, the model generates predicted poverty maps for every region in the sample, including those with no survey coverage. The model implicitly combines and executes two tasks: (1) extracting semantically meaningful observations of, say, housing quality, agricultural productivity, or infrastructure, from raw satellite images; and (2) inferring economic well-being from observing the consumption patterns of these private or public goods (similar to the Engel curve analysis in this study). While the flexibility of the machine learning models helps improve predictive performance, the difficulty in interpretation makes it almost impossible to know or constrain what private or public goods are identified and utilized by the model. Since black-box machine learning models utilize as much information as possible from the input satellite images, it is very likely that the Engel curves of at least some of the observed goods will change (similarly to the tin-roof area variable in this study), introducing biases in the estimated program effects. In this study, we disentangle the two tasks, so that the first task can be framed as a traditional object detection and segmentation task, allowing us to leverage extensive research in computer science; and the second task becomes more transparent, explicit, and the assumptions testable (for example, with Supplementary Figure \ref{fig:engel-tc-diff}).

\section{Discussion}

This paper provides compelling evidence that RCT program evaluations aimed at improving household welfare can be obtained solely based on satellite imagery and deep learning methods. This approach has the advantage of being inexpensive and timely, suggesting great promise as a complement and in some cases as a substitute to in-person survey data collection methods.

However, it bears noting that a fundamental limitation to evaluating programs based on satellite imagery is that in order to be measurable from space, programs being evaluated have to generate impacts on the built landscape. This prevents applicability to programs targeted at addressing development challenges that are unlikely to impact the built environment such as improved teaching methods at schools. Another limitation is that welfare is a household or individual concept whereas satellite images capture characteristics about a place. Mapping household welfare to housing as we do here requires a tight mapping between structures and households through limited mobility. While migration rates are very low in the GiveDirectly study area \cite{egger2019general}, this may be a challenge for programs that  impact mobility, such as transportation infrastructure programs.

\clearpage

\clearpage
\clearpage
\clearpage
\clearpage

\printbibliography
\end{refsection}
\clearpage

\section{Methods}

\paragraph{Constructing the Treatment Intensity Map.}

To construct the treatment intensity map, we utilize data from a baseline census, which was conducted by the authors of the original paper in 2014--2015. The census identified all 65,385 households (roughly 280,000 people) residing in 653 villages in the study area, recorded their GPS coordinates, whether each household was eligible for the GiveDirectly cash transfer, and whether they had been randomized into the treatment or control group \cite{egger2019general}. To address the measurement errors of the GPS collection devices, we discard 58 outliers (living more than 2 kilometers away from the village centers) and impute those and other 4 missing GPS coordinates with village center coordinates. Then, we convert these household records into a raster map. We lay out a regular grid, and count, in each grid cell, the number of households that ultimately received the GiveDirectly cash transfer (see Figure \ref{fig:schematic} and Figure \ref{fig:map}a). Grid cells containing no eligible households are excluded. To account for pre-determined policy intensity differences, we record (and later control for) the number of households that were eligible for the cash transfer, regardless of whether they had been randomized into the treatment or control group.

\paragraph{Obtaining High-resolution Daytime Satellite Images.}

We utilize high-resolution daytime satellite images from Google Static Maps \cite{gsm}. These images have a spatial resolution of about 30cm per pixel (at equator), and contain only the RGB (red, green, blue) bands (see Figure \ref{fig:schematic}d and Supplementary Figure \ref{fig:chips} for examples). These images come from a variety of commercial providers such as Maxar (formerly DigitalGlobe) and Airbus, and have been seamlessly mosaicked together. They have also been geo-referenced and pre-processed to remove clouds and address other data quality issues. Google does not provide the exact timestamps for these images, but we estimate that they were taken in 2019, most likely on Dec 30, 2019. The dates for retrieving these images from the Google Static Maps API are between Feb 19 and Feb 21, 2020, and the Google Earth Pro imagery archive reflects that the closest available images in the study area were from Dec 30, 2019. Multiple other satellite images taken in February, March, July, August and September 2019 are also available in the study area, indicating that the images used in this study are most certainly from 2019.

\paragraph{Extracting Housing Quality Metrics with Mask R-CNN.}

We first leverage a state-of-the-art deep learning model, Mask R-CNN \cite{he2017mask} to segment buildings---that is, to detect each building and the pixels that they occupy---in the Google Static Map satellite images. We then convert the pixel-wise predictions to polygons, and extract housing quality metrics related to the size of the building and the roof materials from each polygon (see Figure \ref{fig:schematic}e and Supplementary Figure \ref{fig:chips} for examples).

Loosely speaking, the Mask R-CNN model operates as follows. First, the model proposes a large number of ``regions of interest'', each of which potentially contains a building. Then, the model uses convolutional filters to identify patterns within the proposed region that are indicative of the presence of buildings, such as the sharp edges, the highly reflective roofs, and the building shadows. Finally, the model predicts whether each proposed region contains a building, as well as whether each pixel is occupied by the building.

We train the Mask R-CNN model with a multi-step process and a transfer learning framework, as described in greater detail in Supplementary Materials \ref{sec:appendix-train}. Publicly available building footprint datasets in rural and low-income regions are rare, and they often differ substantially in spatial resolution, sensor instrument, and landscape from inference images (that is, the target images that the model will make predictions for). Relying solely on publicly available training data is therefore insufficient for achieving satisfactory predictive performance. We curate a set of in-sample annotations by randomly sampling 120 images from all the Google Static Map images in the study area, and manually creating high-quality building footprint annotations for them. We pre-train the Mask R-CNN model on large, publicly available datasets such as COCO (Common Objects in Context) and Open AI Tanzania, and fine-tune them on this set of in-sample annotations.

The model predictions are highly accurate. The overall F1 score (a standard performance metric for instance segmentation) on a random subset of inference images is 0.79 (Supplementary Figure \ref{fig:prcurve}). The F1 score is the harmonic mean of precision (the proportion of model-identified buildings that are actual buildings) and recall (the proportion of actual buildings that are correctly identified by the model). Here, a building is deemed to be correctly identified if the predicted pixel mask and the ground truth pixel mask have sufficient overlap (more precisely, if the intersection of the two masks is more than 50\% of the union of the two masks). As a reference point, the top winner in the 2nd SpaceNet building footprint extraction competition reported an F1 score of 0.69 \cite{spacenet2}. This demonstrates that the Mask R-CNN model used in this study performs well, although building footprint segmentation in rural, less complex scenes is generally easier than in modern cities so these metrics are not directly comparable.

We post-process the model-predicted pixel masks by converting them to polygons, and simplifying the polygons with the Douglas-Peucker algorithm with a pixel tolerance of 3. For each polygon, we compute two housing quality metrics: building footprint and type of roof materials. We then lay out a regular grid, assign each building to grid cells based on the centroids of the polygons, and aggregate to obtain two metrics at the pixel level: building footprint (Figure \ref{fig:map}b) and tin-roof area (Figure \ref{fig:map}c).

First, we measure the size of each building polygon and convert it to square meters. We correct for area distortion, which is induced by the Web Mercator projection system that the Google Static Map uses. This metric may appear larger than what one expects for the size of homes in a low-income context (Figure \ref{fig:engel}), because (1) it represents the footprint of the entire building, which is typically larger than the size of the livable area; and (2) it accounts for both residential and non-residential structures, since the model is not able to distinguish between the two.

Second, we estimate the types of roof materials based on the colors of the roofs, and compute the footprint of tin-roof buildings in each grid cell. For each building, we take all the pixels associated with the given building instance, and assign a ``representative'' roof color by computing the average values in the RGB (Red, Green, Blue) channels. Since the Euclidean distances between color vectors in the RGB color space does not reflect perceptual differences, we project all the RGB color vectors to the CIELAB color space, and cluster these roof color vectors into 8 groups by running the K-means clustering algorithm. We further classify these 8 groups into three types of roof materials: tin roof, thatched roof, and painted roof (Supplementary Figure \ref{fig:colors}), and compute the total footprint of tin-roof buildings.

\paragraph{Obtaining the Night Light Data.}

To measure nighttime luminosity, we use the Visible Infrared Imaging Radiometer Suite (VIIRS) Day/Night Band (DNB) composite images hosted on Google Earth Engine \cite{viirs, elvidge2017viirs}. The VIIRS-DNB data product excludes areas impacted by cloud cover and correct for stray light \cite{mills2013viirs}. However, it has not been filtered to screen out lights from aurora, fires, boats, and other temporal lights, and lights are not separated from background (non-light) values \cite{viirs}. This data product has a native spatial resolution of 15 arc seconds (approximately 463 meters at the equator), and we resample the data by conducting nearest neighbor interpolation when necessary. We average over all the monthly observations in 2019 and construct a single cross sectional observation, to reduce seasonality effects and for consistency with the daytime satellite imagery (Figure \ref{fig:map}d). The VIIRS-DNB data product is considered superior to the more widely used night light data, DMSP-OLS (the United States Air Force Defense Meteorological Satellite Program, Operational Linescan System) because it preserves finer spatial details, has a lower detection limit and displays no saturation on bright lights \cite{elvidge2013viirs}. This ensures that we conduct a fair comparison with the most modern and high-quality night light data product.

\paragraph{Estimating the Program Effects on Housing Quality.}

The main econometric specification for Figure \ref{fig:ate} is as follows
\begin{align}
    y_i = \sum_{k \in K} \tau_k \bm{1}\{x_i = k\} + \sum_{m \in M} \beta_m \bm{1}\{e_i = m\} + \epsilon_i
    \label{eq:ate_bin}
\end{align}
where each observation $i$ represents a $0.001^{\circ} \times 0.001^{\circ}$ grid cell (approximately 100m$\times$100m); $\tau_k$ represents the estimate of interest: the treatment effects of the unconditional cash transfer on remotely sensed outcomes; $x_i$ denotes the number of recipient households per grid cell (equivalent to the amount of cash infusion in \$1000); $e_i$ denotes the number of eligible households per grid cell, with $m \in M = \{0, 1, 2, 3, \cdots\}$; and $y_i$ denotes remotely sensed outcomes: night light, building footprint, and tin-roof area. To account for pre-existing differences in population density or wealth, which may cause non-random variation in treatment intensity, we flexibly control for the number of eligible households per grid cell, and exclude grid cells with no eligible households. Because the grid cells are fairly small and the number of observations for $k>2$ is small, we bin the number of recipient households into four bins $k \in K = \{0, 1, 2, 2+\}$, to preserve statistical power. Standard errors are calculated \`a la Conley, with a uniform kernel and a 3km cutoff \cite{conley1999gmm, conley2008spatial, hsiang2010temperatures, are212conley}. To reduce the effects of outliers (due to sensor malfunctioning or machine learning model prediction errors), we winsorize all remotely sensed variables at the 99 percentile.

We run 100 placebo simulations to further demonstrate the validity of the main specification. In each simulation, we randomly assign half of the 68 groups of villages to the high-saturation group, and the other half to the low-saturation group. In the high-saturation groups, we randomly assign 2/3 of the villages to the treatment group (and the rest to the control group); whereas in the low-saturation group, we assign only 1/3 of the villages to the treatment group (and the rest to the control group). This mimics the two-tier randomization scheme of the original trial \cite{egger2019general}. Using these simulated placebo treatment status variables, we estimate the placebo treatment effects with the econometric specification described in Equation \ref{eq:ate_bin}.

To compute a single pooled treatment effect, we make an assumption of linear treatment effects---every transfer of \$1,000 has an effect of the same magnitude, regardless of the treatment intensity in that geographical area. The resulting econometric specification is as follows
\begin{align}
    y_i = \tau x_i + \sum_{m \in M} \beta_m \bm{1}\{e_i = m\} + \epsilon_i
    \label{eq:ate_pooled}
\end{align}
where $\tau$ is the ``average'' treatment effect, and all else remain the same as in Equation \ref{eq:ate_bin}. We conduct two-sided t-tests to assess statistical significance.

\paragraph{Estimating the Engel Curves.}

An Engel curve describes how household expenditure on a particular good varies with income---a relationship that can be used to infer households' economic well-being from the consumption patterns of a limited subset of goods \cite{elbers2003micro, tarozzi2009using, young2012african, atkin2020new}. The mathematical formulation is
\begin{align}
    Q_{hp} = F_p(W_h) + \epsilon_{hp}
    \label{eq:engel_np}
\end{align}
where household $h$ with $W_h$ wealth (or other measures of economic well-being) would consume $Q_{hp}$ quantities of a normal good $p$, and $F_p(\cdot)$ represents the Engel curve for product $p$ in the population. With a linearity assumption, this can be simplified to be
\begin{align}
    Q_{hp} = \alpha_p + \beta_p W_h + \epsilon_{hp}
    \label{eq:engel_linear}
\end{align}
where $\alpha_p$ is the intercept and $\beta_p$ is the slope of a linear Engel curve.

In this study, we estimate the Engel curves---the relationships between remotely sensed metrics and survey-based measures of economic well-being---based on the endline survey of the original GiveDirectly trial, which includes a representative set of 4,578 geo-coded households who were eligible for the transfer. The households participated in a comprehensive consumption and expenditure survey between May 2016 and June 2017, after the distribution of cash transfers. From the surveys, we observe annualized household consumption expenditure, and asset values. Household consumption expenditure is the annualized sum of total food consumption in the last 7 days, frequent purchases in the last month, and infrequent purchases over the last 12 months. Household assets include housing and non-housing assets, but not land values. Housing asset values are measured as the respondent's self-reported cost to build a home like theirs. Non-housing assets include livestock, transportation (bicycles, motorcycles, and cars), electronics, farm tools, furniture, other home goods, and lending or borrowing from formal or informal sources. We do not study land values because they are difficult to value given thin local markets \cite{egger2019general}.

We perform heuristic matching between the buildings and the household survey GPS coordinates, to link variables in the survey with remotely sensed variables. First, we take the baseline census data, which geo-coded every single household who lived in the study area, and assign every building in the satellite images to its closest census GPS coordinate, if the distance between the two was within 250m. This ensures that every building is matched to at most one household. Second, we match GPS coordinates from the survey with GPS coordinates from the census. While the same household supposedly had the same geo-location, these two often differed because of the measurement errors of the GPS collection devices, and because the coordinates might be recorded anywhere on the participants' plots and not necessarily in their primary residence. We similarly assign each survey GPS coordinate to its closest census GPS coordinate, if the distance between the two was within 250m. In cases of multiple surveys being assigned to the same census coordinate, we keep the closest survey. The final sample contains only census observations that are matched with both buildings in the satellite images and survey records, and consists of 1,904 treatment households and 1,844 control households.

The Engel curves are estimated with only the control group (Figure \ref{fig:engel}a and Supplementary Figure \ref{fig:engel-housing}a, \ref{fig:engel-nonhousing}a and \ref{fig:engel-consumption}a). They are estimated both non-linearly with LOESS (see Equation \ref{eq:engel_np} and the solid lines in Figure \ref{fig:engel}a) and linearly (see Equation \ref{eq:engel_linear} and the dotted lines in Figure \ref{fig:engel}a). When fitting LOESS, we allow for locally-fitted quadratic polynomials, and use 75\% of the data points for each fit. We test for the non-linearity of the Engel curves in a separate procedure. We first run a linear regression, take the residuals, and fit the residuals with a natural (cubic) spline with 5 knots. We then conduct a two-sided F-test on the coefficients of the natural spline basis, and reject the null hypothesis (linearity) if these coefficients are jointly significant. We cannot reject linearity for any of the three proxies in Figure \ref{fig:engel} (building footprint: $F(1,838)=0.37$, $p=0.829$; tin-roof area: $F(1,838)=0.79$, $p=0.533$; night light: $F(1,838)=0.39$, $p=0.814$). To minimize the influence of outliers, we winsorize annual expenditure, housing assets, non-housing assets and total assets at the 1 and 99 percentile of the eligible and non-eligible sample, respectively. We winsorize at the 1 percentile as outliers with a large amount of debt exist and could potentially drive the results otherwise. We similarly winsorize all the remotely sensed variables at the 99 percentile for the eligible and non-eligible sample. We exclude a small number of renters who do not own any housing assets (31 treatment households, 32 control households, and 55 ineligible households), to simplify the interpretation of the Engel curves.

\paragraph{Recovering the Program Effects on Economic Well-being.}

We adapt a prior mathematical formulation that uses the Engel curve to infer changes in economic well-being \cite{young2012african}. Suppose that one is interested in studying the effect of a plausibly exogenous treatment $Z$ on, say, wealth $W$ (denoted $\hat{\tau}_{W}$), but can only inexpensively observe its effect on the consumption of product $p$ (denoted $\hat{\tau}_{Q_p}$). Recall that $\hat{\beta}_{p}$ is the estimated slope of the linear Engel curve in Equation \ref{eq:engel_linear}, then
\begin{align}
    \hat{\tau}_{W} = \hat{\tau}_{Q_p} / \hat{\beta}_{p}
    \label{eq:engel_beta}
\end{align}
Using a formula for propagation of error (or the multivariate Delta method), one can derive the standard error for $\hat{\tau}_{W}$ as follows. This derivation is based on prior work \cite{young2012african}, but additionally accounts for the precision of the slope of the Engel curve.
\begin{align}
    \left(\frac{\hat{\sigma}(\hat{\tau}_{W})}{\hat{\tau}_{W}}\right)^2 = \left(\frac{\hat{\sigma}(\hat{\tau}_{Q_p})}{\hat{\tau}_{Q_p}}\right)^2 + \left(\frac{\hat{\sigma}(\hat{\beta}_{p})}{\hat{\beta}_{p}}\right)^2
    \label{eq:engel_se}
\end{align}

A key assumption of this approach is that $\hat{\beta}_p$ does not depend on $Z$---that is, the Engel curve does not change in direct response to the treatment---also termed the conditional independence assumption \cite{tarozzi2009using}.

We estimate the treatment effects on wealth (or other measures of economic well-being) according to Equation \ref{eq:engel_beta} and Equation \ref{eq:engel_se}, with the treatment effect estimates for remotely sensed variables, and the slopes of the Engel curves. We compare the satellite-derived estimates against the survey-based estimates, taken from Table 1, Column 1 in the original paper \cite{egger2019general}, which were based on the endline household survey data (Figure \ref{fig:engel}b).

\clearpage

\section*{End Notes}

\subsection*{Data and Code Availability}

This paper makes use of restricted access data, which contain personally identifying information of survey participants. Satellite images used in the analyses come from the Google Static Maps API at \url{https://developers.google.com/maps/documentation/maps-static/overview}, and redistribution is not possible. However, de-identified data necessary to reproduce all the figures and statistical analyses are freely available at \url{https://github.com/luna983/beyond-nightlight}.
All the codes are available at \url{https://github.com/luna983/beyond-nightlight}.

\subsection*{Acknowledgements}

We thank Edward Miguel, Jeremy Magruder, Ben Faber, Marshall Burke, Joshua Blumenstock, Supreet Kaur, Ethan Ligon, Elisabeth Sadoulet, Alain De Janvry, Aprajit Mahajan, the participants in the AGU Fall Meeting 2019 (Session GC34C), the UC Berkeley Trade Lunch, Development Workshop, Development Lunch, and Good Data Seminar for feedback.
We thank Edward Miguel, Michael Walker, Dennis Egger, Johannes Haushofer, Paul Niehaus, and the rest of the GiveDirectly team, for generously sharing the dataset with us and responding to our inquiries.

\subsection*{Author Contributions}

L.Y.H. initiated the project, assembled the data, performed the empirical analyses, created the figures, and wrote the paper. L.Y.H., S.H. and M.G.N. collaboratively iterated on the project idea and result interpretation, and edited the paper.

\subsection*{Ethics Declaration}

The GiveDirectly randomized controlled trial and field survey received IRB approval from Maseno University and the University of California, Berkeley. The AEA Trial Registry RCT ID is AEARCTR-0000505. Informed consent was obtained from all human research participants.

The authors declare no conflicts of interest.

\subsection*{Additional Information}

Supplementary Information is available for this paper.
Correspondence and requests for materials should be addressed to Luna Yue Huang (\texttt{yue\_huang@berkeley.edu}).
Reprints and permissions information are available at \url{www.nature.com/reprints}.
\clearpage

\appendix
\begin{refsection}
\renewcommand{\appendixpagename}{Supplementary Information}
\appendixpage

\section{Supplementary Figures}
\renewcommand{\thefigure}{S\arabic{figure}}
\setcounter{figure}{0}

\begin{figure}[!ht]
\centering
\includegraphics[width=5in]{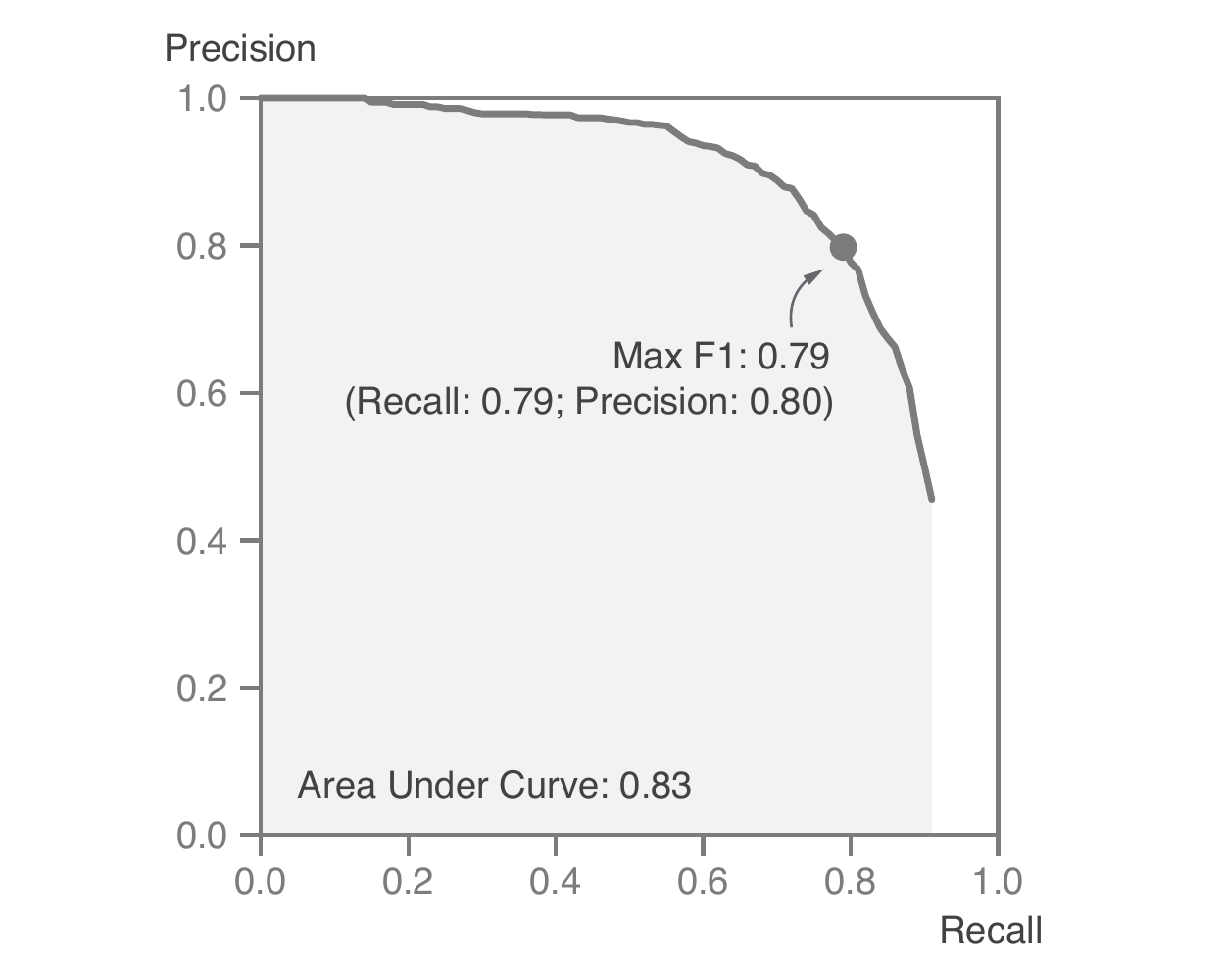}
\caption{
    \textbf{The precision-recall curve of the Mask R-CNN model shows satisfactory predictive performance.}
    The Mask R-CNN model is trained and evaluated with 3-fold cross validation. The evaluation is based on 120 annotated images, which were randomly sampled from all the input satellite images in Siaya, Kenya.
    The Mask R-CNN model outputs a confidence score for every predicted building instance, and the precision-recall curve is generated by varying the confidence score threshold, below which predicted instances are dropped. A higher threshold makes the model more conservative and corresponds to the left portion of the curve (with high precision and low recall), and vice versa. The dot represents the optimal confidence score threshold, obtained by maximizing F1, the harmonic mean of precision and recall. The main model used in this study employs the optimal threshold, and has a recall of 0.79 and a precision of 0.80.
}
\label{fig:prcurve}
\end{figure}
\clearpage
\begin{figure}[!ht]
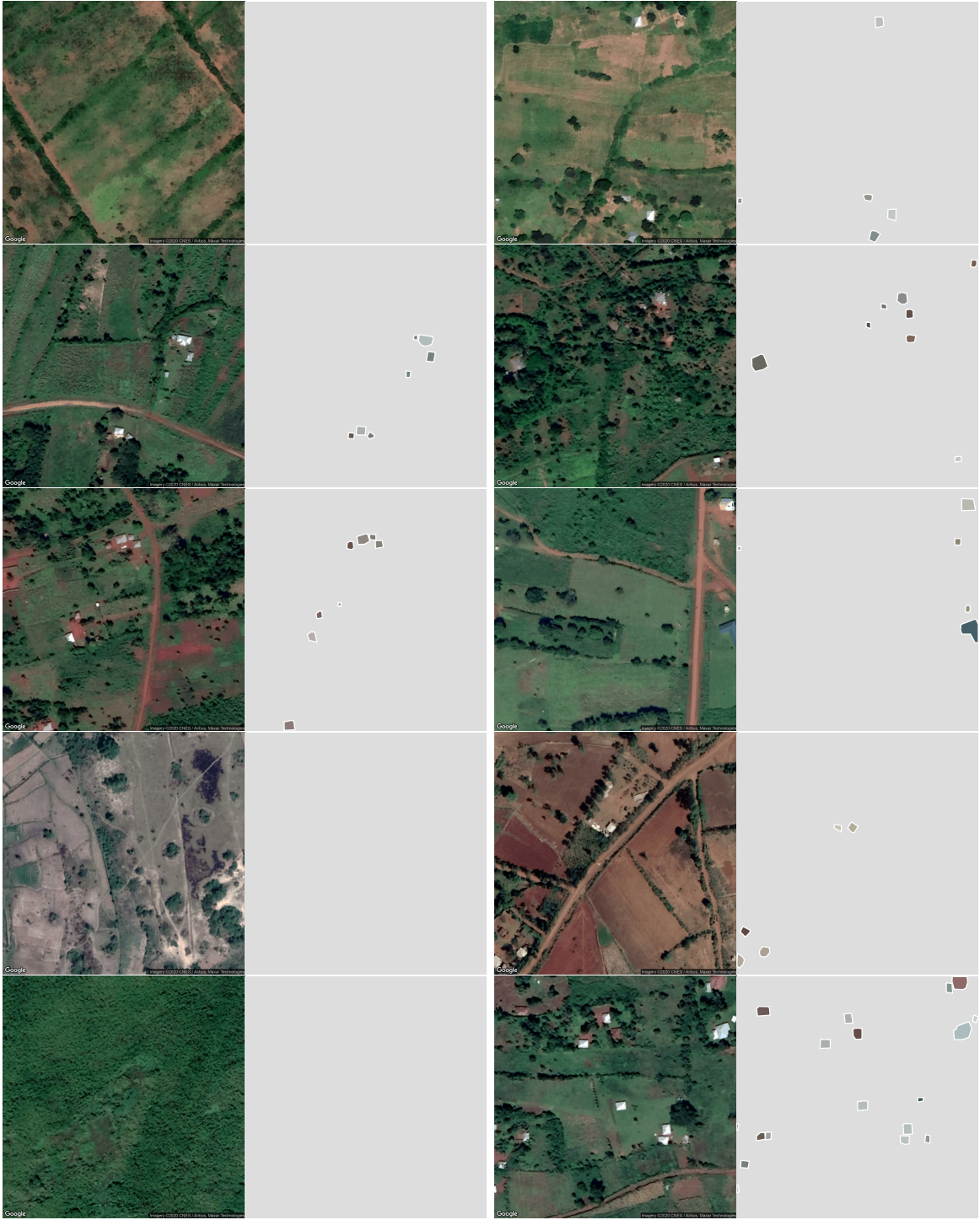

\centering
\newcounter{i}
\forloop{i}{0}{\value{i} < 10}{
    \includegraphics[width=1.5in]{figures/fig-schematic/fig-chips-img\arabic{i}.png}%
    \includegraphics[width=1.5in]{figures/fig-schematic/fig-chips-poly\arabic{i}.pdf}%
}
\caption{
    \textbf{Ten randomly sampled pairs of input images and deep learning predictions.}
    Ten images are randomly sampled from all the input satellite images in the GiveDirectly study area.
    Each predicted building is outlined in white and filled with the ``representative'' roof color.
}
\label{fig:chips}
\end{figure}
\clearpage
\begin{figure}[!ht]
\centering
\includegraphics[width=6in]{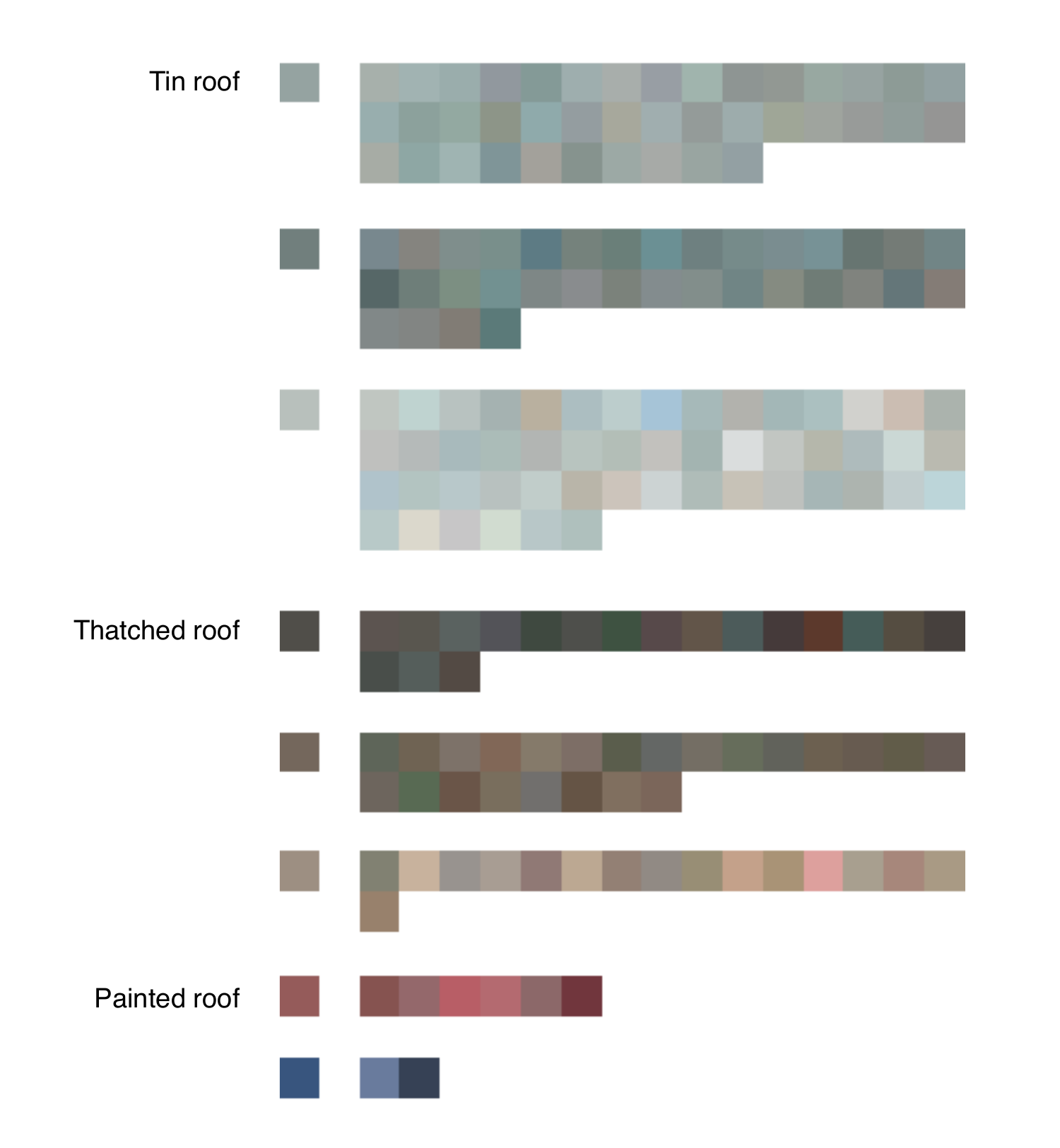}
\caption{
    \textbf{The distribution and grouping of roof colors.}
    All the buildings in the GiveDirectly study area are split into eight groups by a K-means clustering algorithm, based on their roof colors. The color block on the left represents the ``average'' roof color of the cluster, and the color blocks on the right represent a random subset of all the roof colors in the given cluster. The number of color blocks on the right is proportional to the size of the cluster. The eight groups are further grouped into tin roof, thatched roof, and painted roof.
}
\label{fig:colors}
\end{figure}
\clearpage
\begin{figure}[!ht]
\centering
\includegraphics[width=6in]{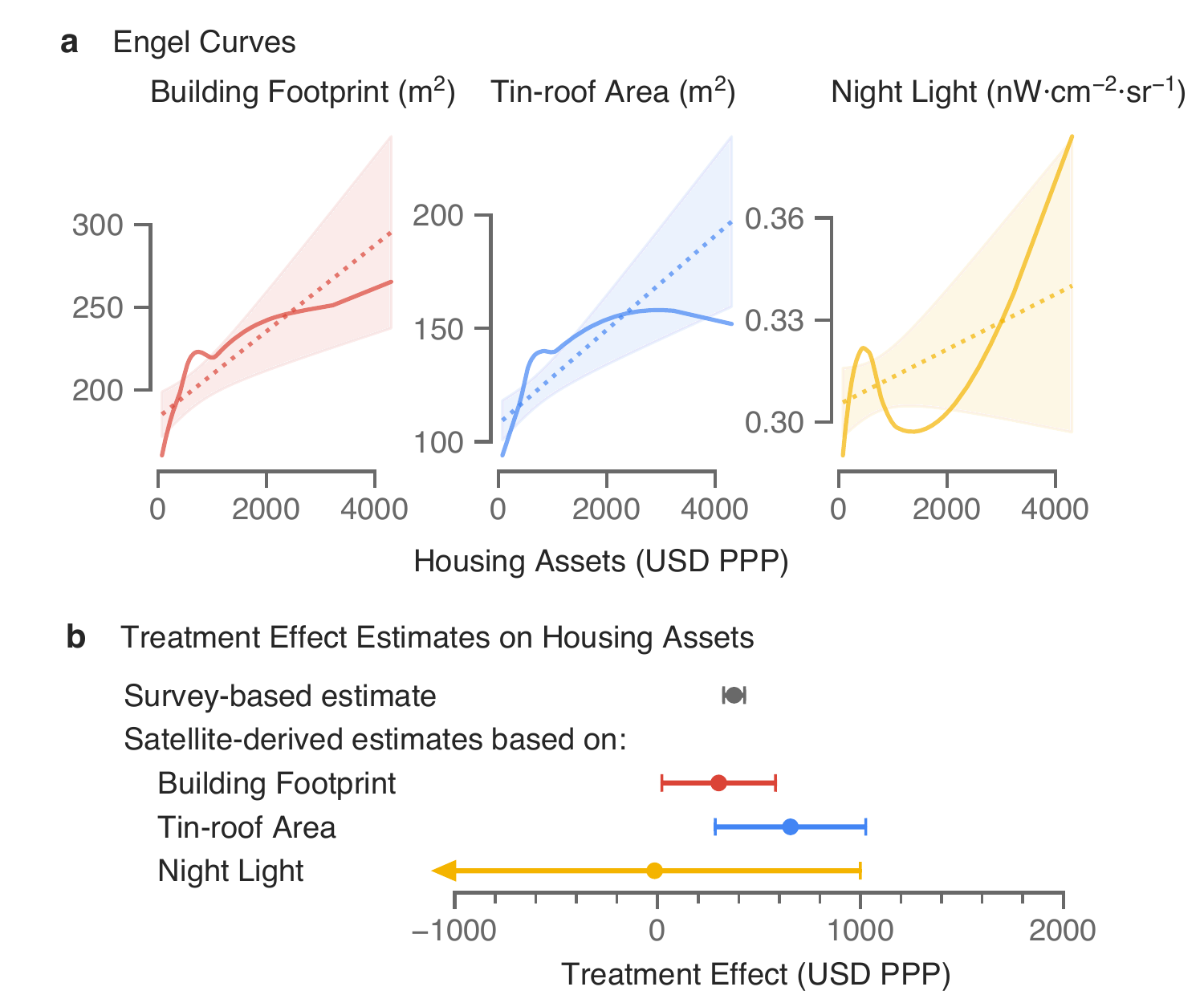}
\caption{
    \textbf{The treatment effect on housing assets can be similarly recovered by scaling the effect on building footprint.}
    \textbf{a} The Engel curves of building footprint, tin-roof area, and night light, estimated with LOESS (solid line) or a linear regression (dotted line). The shaded regions represent the 95\% confidence intervals for the latter.
    \textbf{b} Comparing the survey-based versus satellite-derived treatment effects. The dots show the point estimates. The error bars show the 95\% confidence intervals, with the arrow(s) marking upper/lower bounds that are out of range (if any).
    $n=1,844$.
}
\label{fig:engel-housing}
\end{figure}
\clearpage
\begin{figure}[!ht]
\centering
\includegraphics[width=6in]{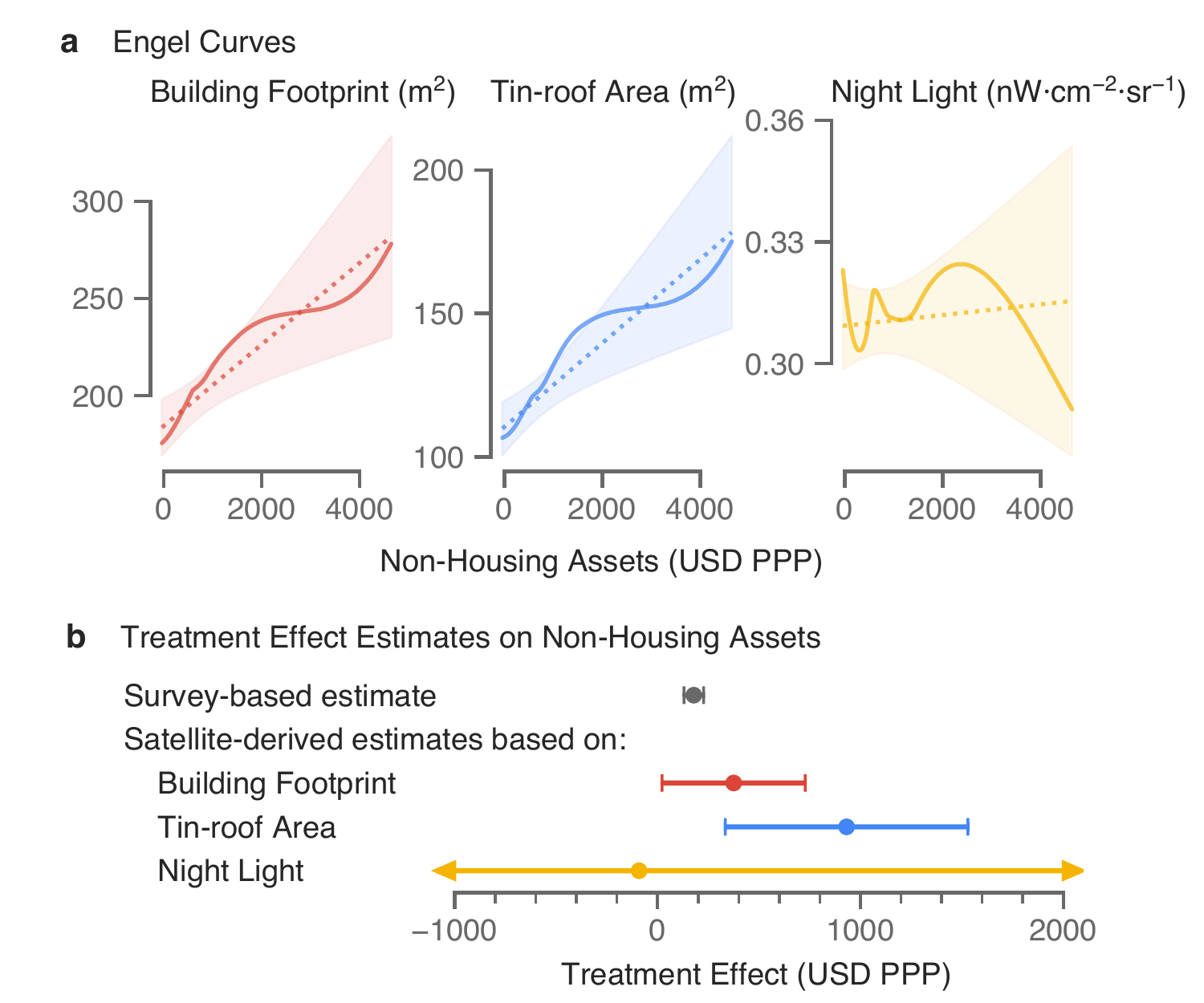}
\caption{
    \textbf{The treatment effect on non-housing assets can be similarly recovered by scaling the effect on building footprint.}
    \textbf{a} The Engel curves of building footprint, tin-roof area, and night light, estimated with LOESS (solid line) or a linear regression (dotted line). The shaded regions represent the 95\% confidence intervals for the latter.
    \textbf{b} Comparing the survey-based versus satellite-derived treatment effects. The dots show the point estimates. The error bars show the 95\% confidence intervals, with the arrow(s) marking upper/lower bounds that are out of range (if any).
    $n=1,844$.
}
\label{fig:engel-nonhousing}
\end{figure}
\clearpage
\begin{figure}[!ht]
\centering
\includegraphics[width=6in]{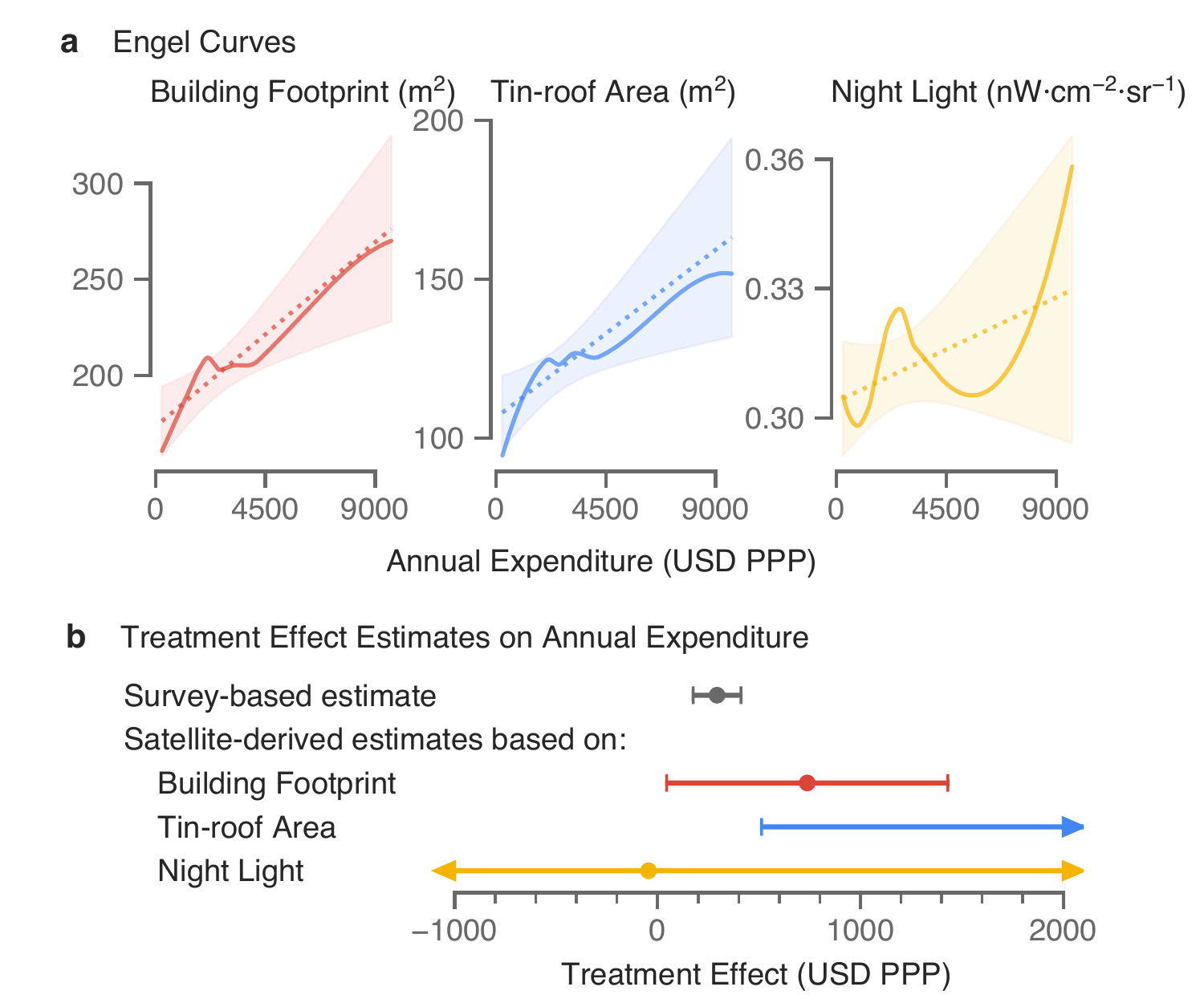}
\caption{
    \textbf{The treatment effect on annual expenditure can be similarly recovered by scaling the effect on building footprint.}
    \textbf{a} The Engel curves of building footprint, tin-roof area, and night light, estimated with LOESS (solid line) or a linear regression (dotted line). The shaded regions represent the 95\% confidence intervals for the latter.
    \textbf{b} Comparing the survey-based versus satellite-derived treatment effects. The dots show the point estimates. The error bars show the 95\% confidence intervals, with the arrow(s) marking upper/lower bounds that are out of range (if any).
    $n=1,843$.
}
\label{fig:engel-consumption}
\end{figure}
\clearpage
\begin{figure}[!ht]
\centering
\includegraphics[width=6.5in]{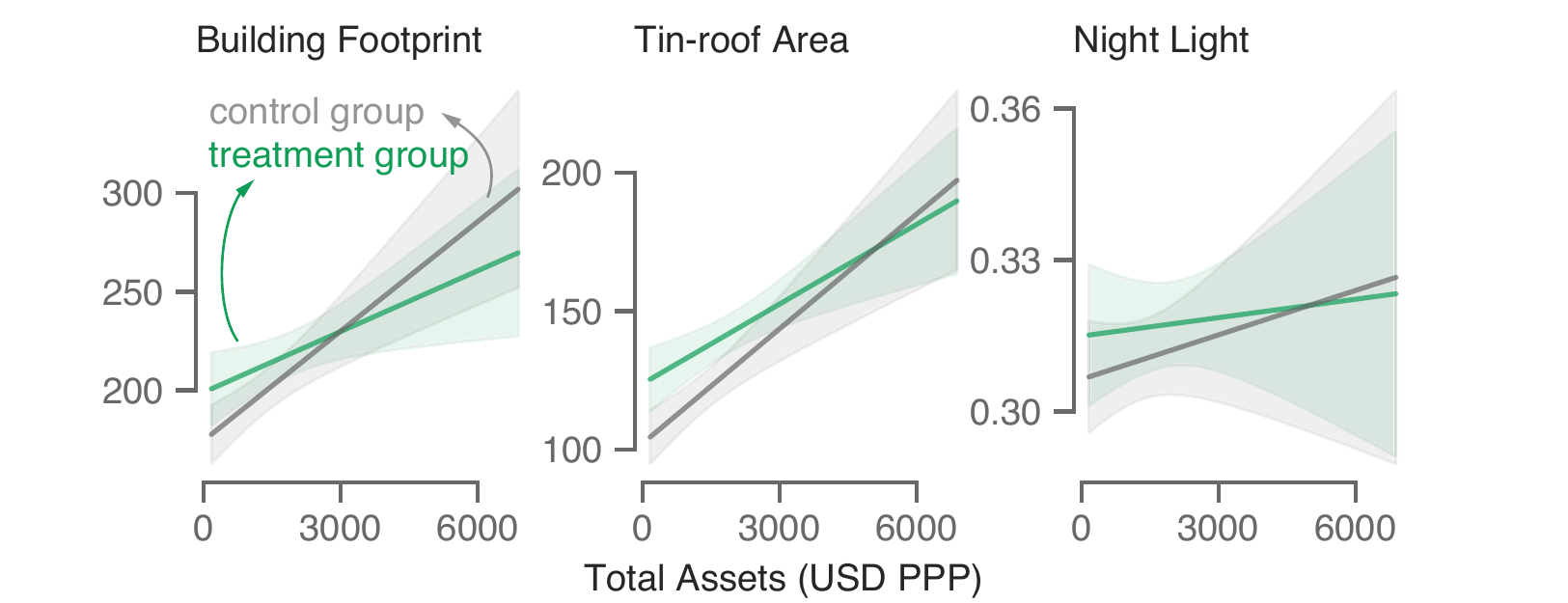}
\caption{
    \textbf{The Engel curves for tin-roof area shifted in response to the cash transfer.}
    The Engel curves for the treatment households (in green, $n=1,904$) and the control households (in gray, $n=1,844$). The shaded regions represent the 95\% confidence intervals.
}
\label{fig:engel-tc-diff}
\end{figure}
\clearpage
\begin{figure}[!ht]
\centering
\includegraphics[width=6.5in]{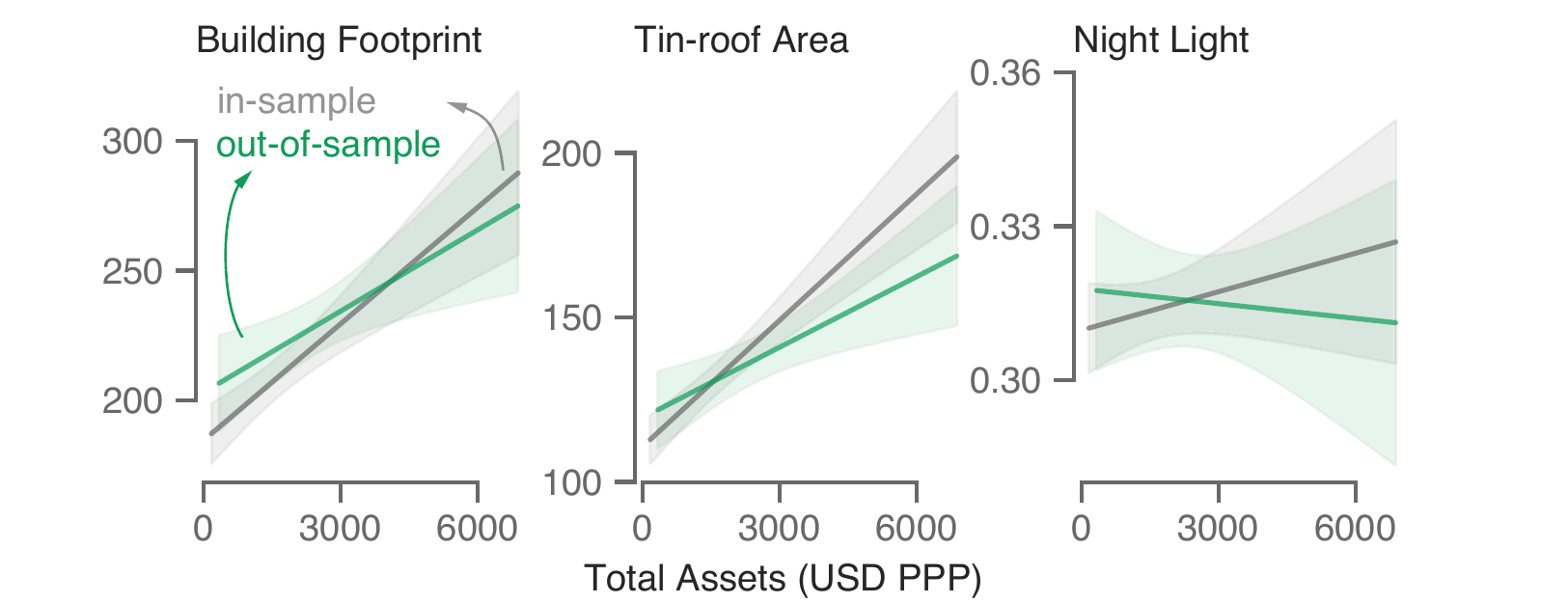}
\caption{
    \textbf{The Engel curves estimated based on in-sample and out-of-sample data are broadly similar.}
    The Engel curves for the in-sample eligible households (in gray, $n=3,748$) and the out-of-sample ineligible households (in green, $n=1,821$) in the GiveDirectly study area. The shaded regions represent the 95\% confidence intervals.
}
\label{fig:engel-ei-diff}
\end{figure}
\clearpage
\begin{figure}[!ht]
\centering
\includegraphics[width=6in]{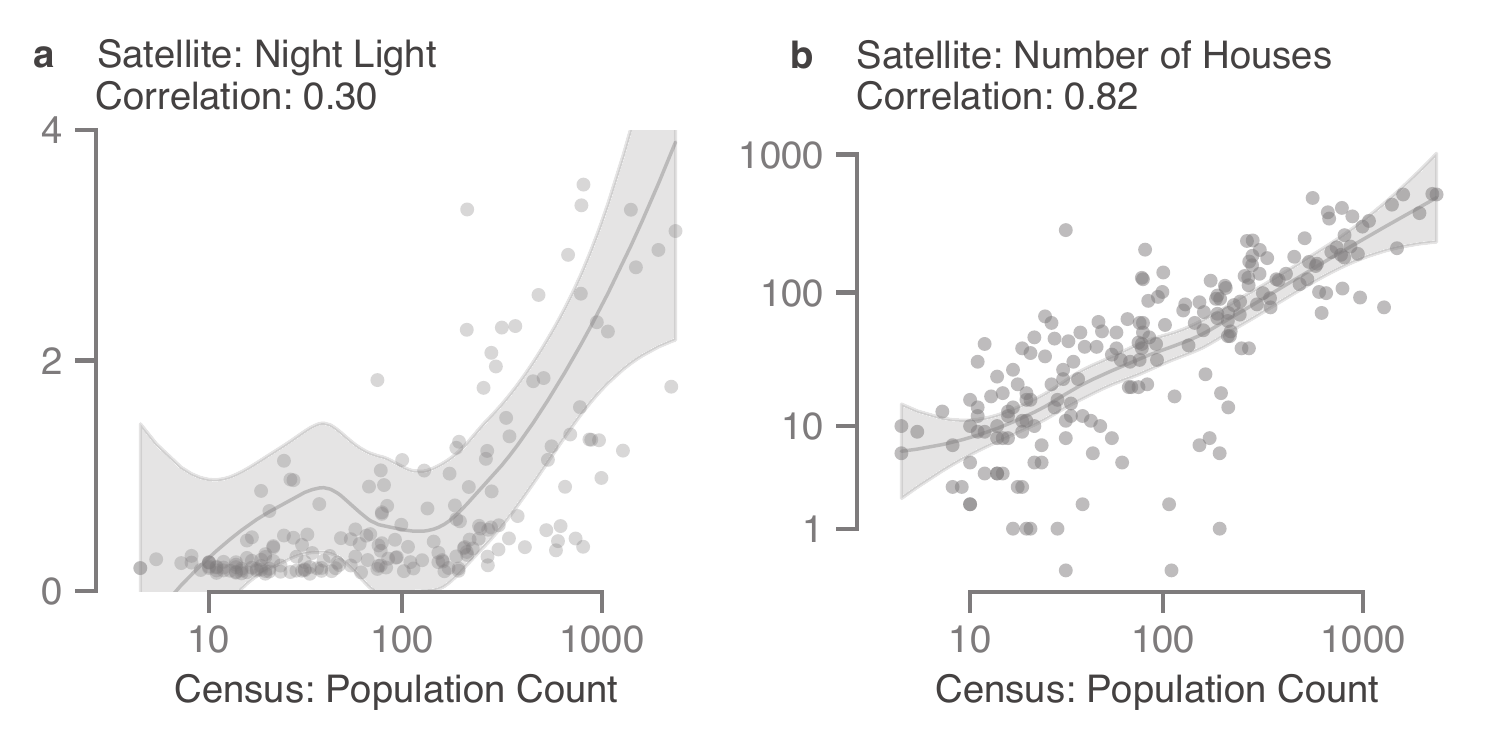}
\caption{
    \textbf{Population count in Mexican villages is more strongly correlated with the number of houses in satellite imagery, compared to night light.}
    The population count is shown in log scale. Each point corresponds to a randomly sampled rural locality in Mexico. Gray lines are estimated LOESS curves, and the shaded regions are the 95\% confidence intervals. The (Pearson) correlation coefficients are reported in the panel subtitles. $n=197$.
}
\label{fig:mx}
\end{figure}
\clearpage

\section{Cost Estimation} \label{sec:appendix-cost}

We estimate that our evaluation approach costs \$0.006 per household, when accounting for imagery acquisition and computing costs. The Google Static Maps API charges users \$0.002 per image request \cite{gsm_billing}. We estimate that our computing cost is roughly \$0.004 per household. Our entire data pipeline can be run within 72 hours on an NC24 instance with 4 K80 GPUs on Microsoft Azure, which costs \$3.60 per hour, and we have analyzed over 60,000 households. This is a liberal estimate that accounts for image downloading, model training, model inference, model validation, and regression analysis. Notably, we do not include labor costs for research and development, as these only need to be incurred once, are not relevant for application of the method, and that such labor costs are difficult to quantify.
\clearpage

\section{Training the Deep Learning Model} \label{sec:appendix-train}

\subsection{Creating In-sample Building Footprint Annotations} \label{sec:appendix-annotate}

We create in-sample building footprint annotations to train the model, and to objectively and quantitatively evaluate model performance. Among the 71,012 satellite images that cover all of the Siaya county in Kenya, we randomly sample 120 images for annotation. We use the Supervisely image annotation web platform to create annotations. On any given image, we outline the boundaries of all the instances of buildings on the image. Buildings that border each other are annotated as separate instances, if there are reasons to believe that they are separate structures (e.g., if they appear to use different roof materials). Half-finished buildings are annotated, although they are fairly rare in the analysis sample.

Some measurement errors can arise from the annotation process, which may in turn impact the predictions of the deep learning model. First, the Google Static Maps logo blocks 1.05\% of the total area of any given image, and structures covered by the logos are not annotated. Second, only the visible parts of the buildings are annotated, but a very small part of some buildings may be partially occluded by trees. Third, the annotation accuracy (and thus potentially prediction accuracy) may be different across buildings with different roof materials. In particular, thatched-roof houses tend to be harder to identify for human annotators than metal-roof houses, because they are typically smaller, not as reflective, and may resemble trees in the overhead imagery.

\subsection{Training the Mask R-CNN Model} \label{sec:appendix-schedule}

We use the Mask R-CNN model \cite{he2017mask} for instance segmentation of buildings on satellite images. The backbone architecture used is ResNet50 with the Feature Pyramid Networks. The model is trained with a learning rate of $5 \times 10^{-4}$ and a batch size of 10. Optimization is conducted with the Adam optimizer. We implement the deep learning pipeline with Python and PyTorch. In particular, we use the official Torchvision implementation of Mask R-CNN. We train the Mask R-CNN model in a transfer learning framework, with a multi-step process as follows.

\paragraph{1. COCO (Common Objects in Context)} The model is first pre-trained with the COCO (Common Objects in Context) data set, a large-scale natural image data set containing 80 object categories and around 1.5 million object instances \cite{coco}. Despite the fact that input images and object categories in COCO are different from target satellite images, pre-training the model with a large-scale dataset often provides meaningful performance gains, even when the model is later transferred across domains.

\paragraph{2. Open AI Tanzania} The model is then fine-tuned on the Open AI Tanzania building footprint segmentation data set, a collection of high-resolution aerial imagery collected by consumer drones in Zanzibar, Tanzania \cite{openaitanzania}. These images are representative of the rural or peri-urban scenes in a developing country context, in terms of the distribution of the density, sizes and heights of the buildings. All the buildings in the drone images are identified, outlined and classified into three categories (completed building, unfinished building, and foundation) by human annotators. This somewhat unusual categorization is due to the fact that there are a large number of unfinished structures in Zanzibar. Most input satellite images in this study contain very few unfinished structures, so we collapse the first two categories into one and drop the third category. The native resolution of the drone images is 7cm, and we down-sample the images to about 30cm to match with the resolution of the target satellite images.

In training time, 90\% of the data are used for training, and the remaining 10\% for validation. In order to guard against overfitting, and choose the best model, in each epoch, we evaluate the performance of the model with the validation set, using average precision with an Intersection over Union (IoU) cutoff of 0.5 as the main evaluation metric. The model is trained for 50 epochs, and the best model (at epoch 43) is saved and loaded in subsequent steps.

\paragraph{3. Supplementary Annotations in Mexico, Tanzania and Kenya} The model is then fine-tuned on a set of 587 annotated high-resolution satellite images from Mexico, Tanzania, and Kenya. The Mexico dataset consists of 199 satellite images corresponding to 8 randomly sampled rural localities studied in Supplementary Figure \ref{fig:mx}. Some of these are historical images with lower data quality and more cloud coverage. These images are pooled and randomly split into a training set (90\%) and a validation set (10\%). The model is trained for 25 epochs, and achieves the best performance at epoch 17.

\paragraph{4. In-sample Annotations} Finally, the model is fine-tuned on a set of 120 in-sample annotated images in Siaya, Kenya (see Section \ref{sec:appendix-annotate} for details). This ensures that training images and inference images belong to the same data distribution. The model is trained on 90\% of the images for 25 epochs, and evaluated with the 10\% held out set. We keep the best-performing model (at epoch 15). This is the main model used for conducting inference on input satellite images in the GiveDirectly study area.

\vspace{1em}\noindent Throughout the training process, we conduct extensive data augmentation to increase the transferability of the model from one dataset to another. We randomly flip the training images horizontally and vertically, randomly jitter the brightness, contrast, saturation, and hue of the images. For the Open AI Tanzania dataset, we also randomly blur and crop the images.

\clearpage

\section{Validation in Mexico} \label{sec:appendix-mx}

\subsection{Results}

We provide additional validation results in rural Mexico, using the 2010 Population and Housing Census \cite{cpv2010}. Population count in a rural village (as reported in the 2010 census), is highly correlated with the number of houses in that village (as identified by the deep learning model), with a Pearson correlation coefficient of 0.82 (Supplementary Figure \ref{fig:mx}b). Population count, however, is only modestly correlated with night light (Supplementary Figure \ref{fig:mx}a). Night light is less sensitive in smaller, less populated villages, a finding that is consistent with prior work \cite{jean2016combining}.

\subsection{Methods}

This comparison is based on the locality-level data set, Principales Resultados por Localidad, or ITER. (A locality is equivalent to a village in rural areas.) To form the analysis sample, we drop all urban localities (defined as having more than 2,500 residents), small localities where the relevant asset measures are masked in the census to protect privacy, and localities where these measures are missing. To avoid covering neighboring urban or rural localities in the satellite images, we exclude rural localities that are closer than 0.01 degree (1.1 km) from other rural localities, or 0.1 degree (11.1 km) from urban localities. Finally, to reduce computation, we randomly sample 200 rural localities, and drop 3 of them, for which Google Static Maps does not have satellite image coverage for.

In the census, each rural locality is geo-coded as a point. Most of the rural localities are small, isolated and surrounded by vegetation or open space, making it feasible to match census records to corresponding satellite images. For each locality, we obtain satellite images that cover an area of roughly $1 \times 1$ km, with the locality coordinate at the center. The images are retrieved from the Google Static Maps API on October 10, 2019, and are likely taken several years after the census. We generate deep learning predictions on these images with the method described in Methods and Supplementary Materials \ref{sec:appendix-train}, but only train the model for the first three steps in Supplementary Materials \ref{sec:appendix-schedule}. For the comparison, we count the number of houses in a locality in the deep learning predictions, and extract the population count variable from the census. Additionally, we download night light data, the Visible Infrared Imaging Radiometer Suite (VIIRS) Day/Night Band (DNB) composite images from 2019.
\clearpage

\printbibliography
\end{refsection}

\end{document}